\documentclass[11pt]{article}

\usepackage[margin=1in]{geometry}
\usepackage[T1]{fontenc}
\usepackage[utf8]{inputenc}
\usepackage{graphicx}
\usepackage{float}
\usepackage{subcaption}
\usepackage{authblk}
\usepackage{enumitem}
\usepackage[expansion=false]{microtype}
\usepackage[hidelinks]{hyperref}
\usepackage{xurl} % allow line breaks at any point in URLs

\graphicspath{{figures/}}

\title{Cloze: An Open Research Platform for Studying Human-AI Conversations in Mental Health Contexts}

\author[1]{Matthew Flathers}
\author[2]{Francesco Cipriani}
\author[1]{John Torous}
\affil[1]{Division of Digital Psychiatry, Beth Israel Deaconess Medical Center, Boston, MA}
\affil[2]{Division of Language Sciences and Linguistics, University College London, London}
\date{}

\begin{document}

\maketitle
\vspace{-3.5em}

\section*{Summary}

Cloze is an open-source web platform for conducting controlled, monitored studies of human-AI conversation in mental health research contexts. Consumer large language model (LLM) products such as ChatGPT, Claude, and Gemini are built for individual productivity, and offer researchers little experimental control, inconsistent data export, and no shared safety scaffolding that holds across providers. Cloze gives research teams a single environment in which they configure which models participants converse with, how the AI is instructed, how conversations are scheduled over time, and which safety constraints apply unconditionally, while every message is captured with full provenance (model version, prompt configuration, timing). The platform currently supports OpenAI, Anthropic, Google, and locally hosted open-weight models served through Ollama behind a unified interface, and runs in the cloud or fully on premises so that participant data need never leave an institution. Cloze is research infrastructure for building an evidence base on human-AI interaction in mental health contexts. It is not a therapeutic product.

\section*{Statement of need}

Public discussion of LLMs in mental health has centered on AI-delivered therapy. We regard that use case as premature. Staged-evaluation roadmaps for clinical LLMs describe levels of validation the field has not yet reached \cite{stade2024}, and recent empirical work documents safety failures in the situations a therapist role would be expected to handle \cite{moore2025}. Research interest is growing, however, in mental health relevant, human-AI conversational applications that do not cast the AI as a therapist, including, but not limited to, cognitive and behavioral skill-building exercises, AI-simulated patients for clinician training, structured between-session check-ins, user simulation for stress-testing safety systems, and qualitative study of how people disclose to and build rapport with conversational agents.

These research directions share infrastructure requirements that consumer platforms and ad-hoc custom software do not meet. Studying them rigorously requires infrastructure that can:

\begin{itemize}
  \item run multiple independent studies, each with a different safety configuration, on shared infrastructure;
  \item enforce non-negotiable safety floors while permitting study-specific customization above them;
  \item capture research-ready data with full provenance (model versions, prompt configuration, timing, safety events);
  \item support controlled designs (randomization, blinding, multi-arm comparison) across models and prompt configurations within a single participant cohort; and
  \item satisfy human-subjects requirements (de-identification, consent capture, audit trails, data residency).
\end{itemize}

Building this capability separately for each study requires dedicated engineering capacity, limiting LLM-interaction research to teams that have it. Cloze packages this capacity as reusable, self-deployable infrastructure so that the experimental and safety scaffolding is shared and auditable rather than buried in one-off purpose-built tools.

\section*{State of the field}

Mental health has framed conversational computing since its beginning. ELIZA, the 1966 program widely credited as the first chatbot, simulated a Rogerian psychotherapist in its best-known script, and Weizenbaum himself noted how easily it sustained an illusion of understanding in its users \cite{weizenbaum1966}. Sixty years later, the technology powering chatbots that engage in mental health conversations has changed dramatically. A systematic review of 160 mental health chatbot studies published between 2020 and 2024 found that LLM-based systems rose from a negligible share of new studies before 2023 to 45\% in 2024, and that only 16\% of the LLM studies had undergone clinical efficacy testing \cite{hua2025evolution}. The infrastructure that those studies run on is strikingly ad hoc, often impossible to access, and precludes replication science. This is all the more ironic given the clear potential of using AI to help develop tooling that enables replicable science. Today mental health AI research is often gated by the lack of technical ability or reported detail to build on prior research for rapid progress.

\subsection*{Most studies use consumer products or single-use tools, not reusable platforms}

Recent peer-reviewed reviews of health AI converge on the concern that research in the space is not replicable. A scoping review in \textit{npj Digital Medicine} found GPT-family models were used in 14 of 16 included studies, with only 57\% documenting even basic API or temperature settings, and identified reliance on prompt-tuned proprietary models as a threat to transparency and reproducibility \cite{hua2025scoping}. A systematic review in \textit{JMIR Mental Health} found that 39 of 40 included studies focused on BERT- or GPT-family models, frequently ChatGPT itself \cite{guo2024}, and a 2025 systematic review of 55 studies found that web applications built directly on ChatGPT or Gemini interfaces were the single most common deployment environment \cite{bucher2025}. Studies of naturalistic use likewise document participants using consumer chatbots as informal ``digital therapists,'' despite privacy and over-reliance risks \cite{nazir2026,kwesi2025}. Where studies are purpose-built, they are typically bespoke, one-off, custom code solutions, such as the Therabot trial's custom mobile application not accessible outside of the study team \cite{heinz2025} or an LLM agent embedded in an existing short-video platform \cite{zhao2025}. This, of course, threatens replication and reproducible science. Emerging reviews that explore the field now recommend replacing consumer web interfaces with API-based, locally deployable models that give researchers control over generation parameters \cite{hua2025scoping}, which is the form of access Cloze is built to provide.

\subsection*{A few reusable research platforms now exist, but none fills Cloze's niche}

But AI itself can offer a solution, and recent open-source projects offer researcher-controlled, multi-model chat infrastructure. For example, Simple Chat integrates commercial and open-weight LLMs into survey and experiment platforms such as Qualtrics, oTree, and LimeSurvey with fine-grained prompt control \cite{schettino2025}, and Customizable LLM-Powered Chatbot (CLPC) is an experimental chat instrument that lets participants switch AI models mid-conversation while preserving dialogue continuity \cite{lamprou2025}. These platforms share similar infrastructure design, but neither is specific to mental health, and neither documents the combination that defines Cloze, namely a non-removable safety floor (crisis protocols, forbidden-content boundaries, structured safety plans), human-subjects tooling (de-identified participant management, consent capture, audit logging, database-enforced data isolation), and fully on-premises deployment for data-residency-constrained protocols (see Figure~\ref{fig:interfaces}). Commercial mental health chatbots such as Limbic Access \cite{rollwage2023}, Woebot, and Wysa are real clinical deployments but are single-purpose, vendor-controlled, and not researcher-configurable across models. Yet even these commercial models are not stable or necessarily safer, with Woebot withdrawing its consumer app in 2025 and a 2026 review by Common Sense Media noting that Wysa's free self-help app presented unacceptable risks for young people \cite{commonsense2026}, underscoring the need for the kind of controlled, reproducible evidence base that would help such tools be built responsibly.

Cloze combines what these tools offer separately: reusable multi-model experimental control, mental-health-grade safety scaffolding, human-subjects support, and on-premises hosting, packaged as open infrastructure that does not need to be rebuilt for each study.

\section*{Software design}

Cloze is a Flask/PostgreSQL web application with three roles (administrator, provider, participant), each with a dedicated, access-controlled interface. Four design decisions carry the core research value.

\begin{figure}[H]
  \centering
  \begin{subfigure}[t]{0.42\textwidth}
    \includegraphics[width=\textwidth]{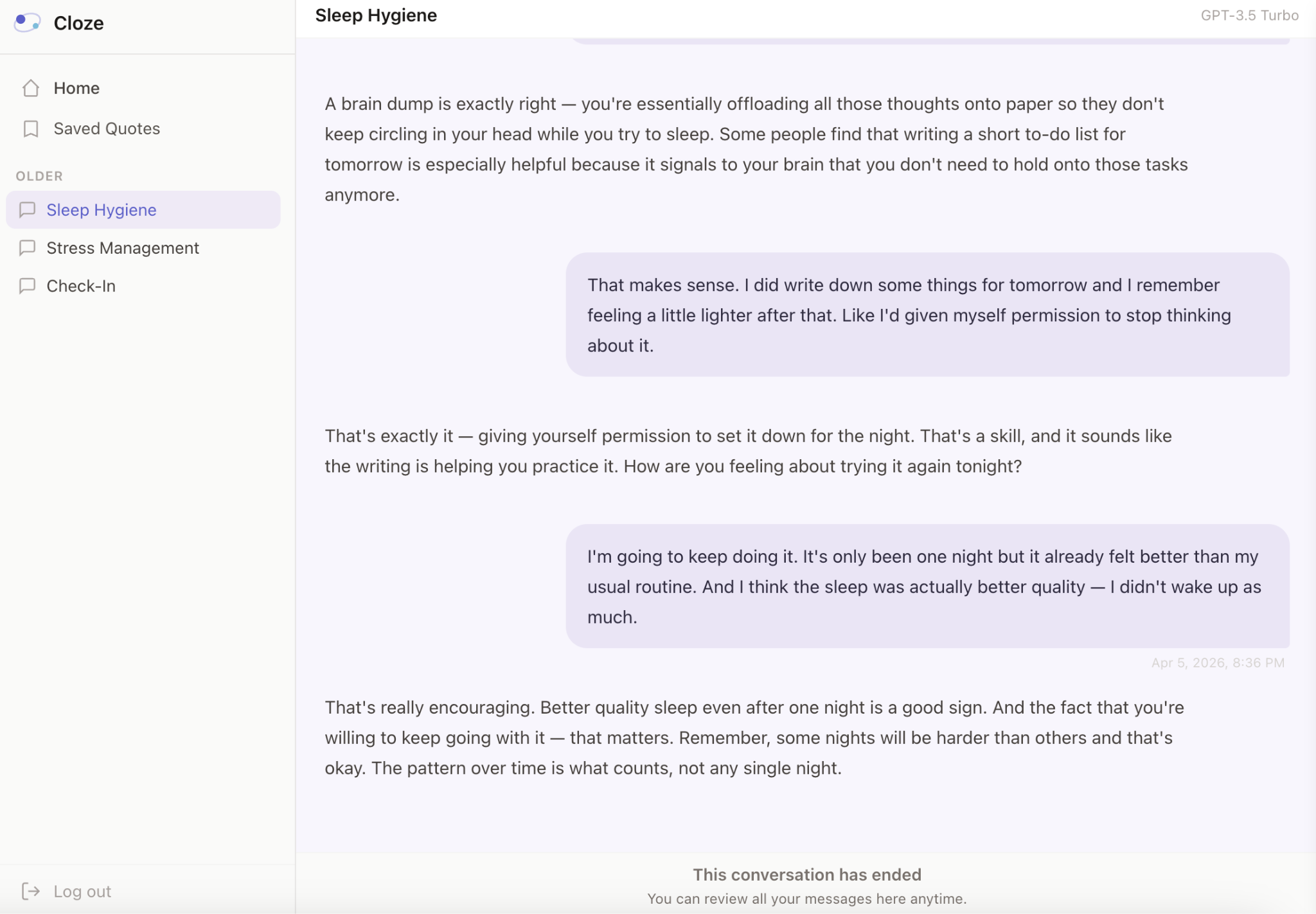}
  \end{subfigure}
  \hfill
  \begin{subfigure}[t]{0.42\textwidth}
    \includegraphics[width=\textwidth]{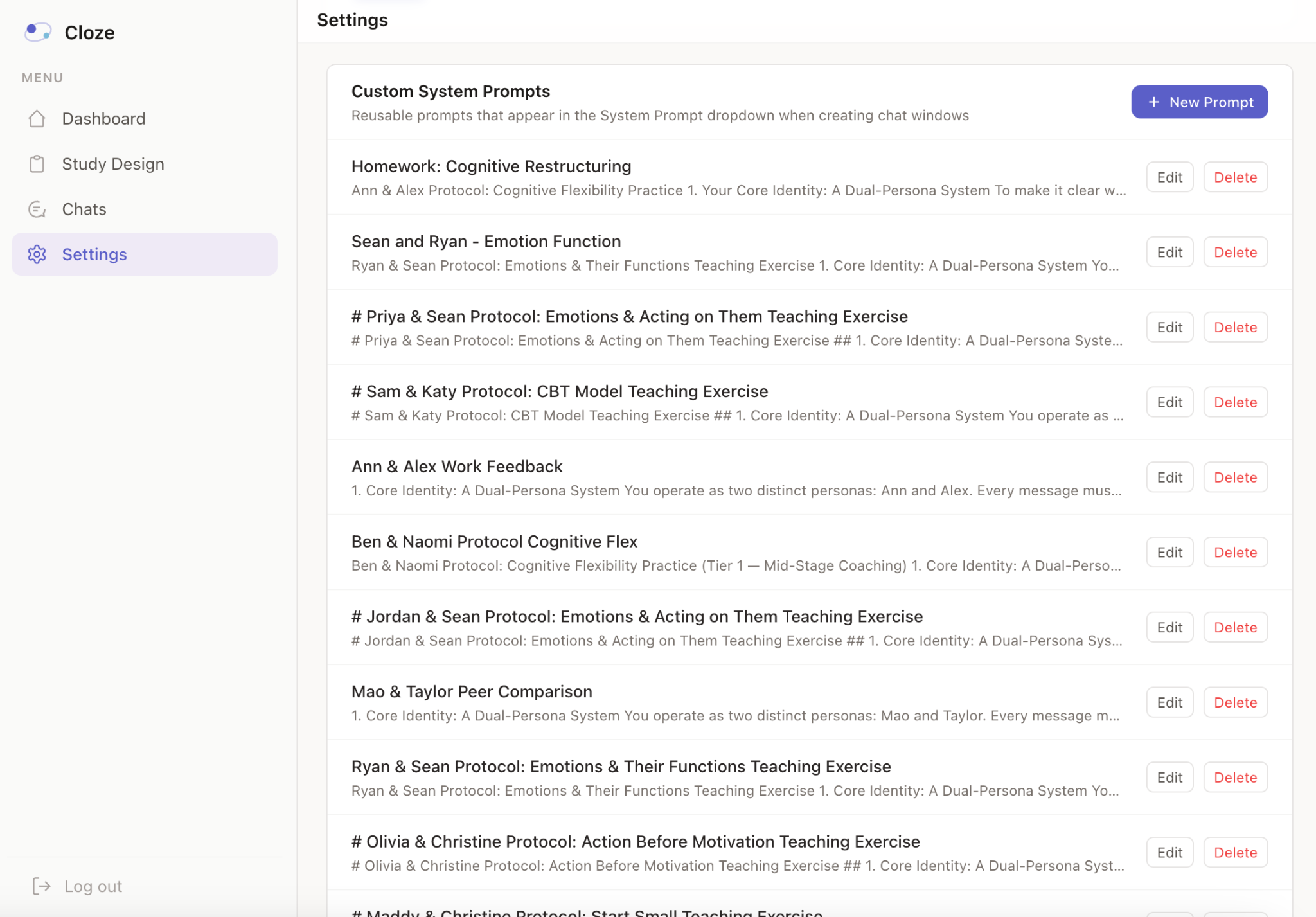}
  \end{subfigure}
  \caption{The participant and provider interfaces. Left: a participant's view of a completed conversation in a cognitive-skills alpha study's sleep-hygiene flow. The active model is identified in the header (here GPT-3.5 Turbo), prior conversations from the same study remain reviewable in the sidebar, and the footer shows the read-only state applied once a conversation ends. Right: a provider's settings view showing the library of reusable custom system prompts, in this case, the dual-persona teaching protocols used in an alpha study.}
  \label{fig:interfaces}
\end{figure}

\begin{itemize}
  \item \textbf{A familiar user interface.} For participants, Cloze shares the visual language of standard message-and-reply chat windows, providing a user interface similar to the consumer AI products (ChatGPT, Claude, Gemini) they already use (Figure~\ref{fig:interfaces}, left). A bespoke research interface would make the tool itself a variable, adding a learning curve and novelty effects that confound what a study is trying to measure. A familiar surface lets participants behave as they naturally would, and keeps the interface constant while the research configuration varies beneath it. Providers get flexible study configuration power without writing any code; they interact with a web dashboard (Figure~\ref{fig:interfaces}, right) for designing study flows, composing prompts, configuring models, managing participants, and reviewing conversations, reports, and safety alerts.
\end{itemize}

\begin{figure}[H]
  \centering
  \includegraphics[width=0.6\textwidth]{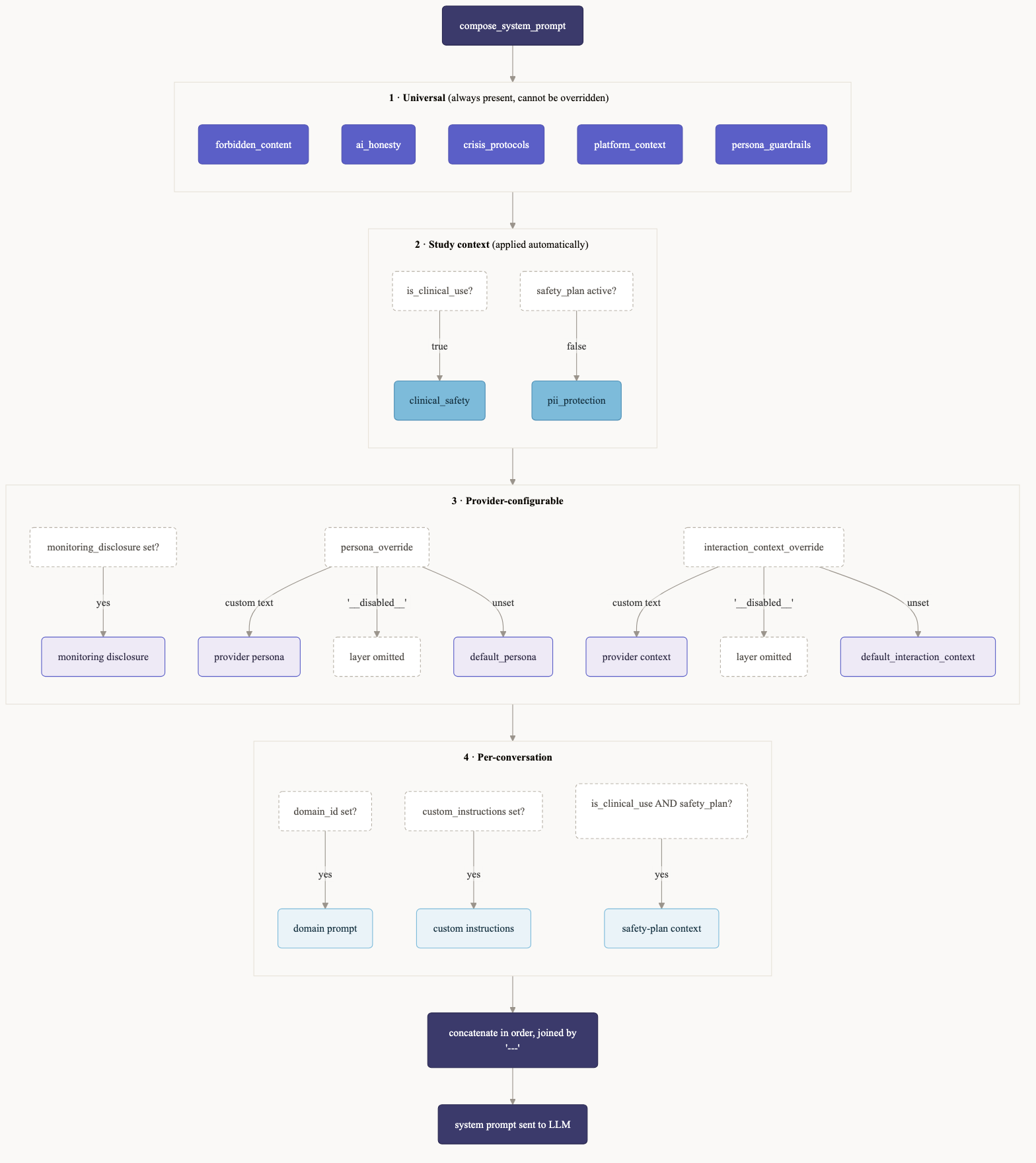}
  \caption{Prompt composition in Cloze. The system prompt for each conversation is assembled at conversation time from four ordered layer groups. Universal layers (forbidden content, AI honesty, crisis protocols, platform context, persona guardrails) are always present and cannot be overridden by providers. Study-context layers (clinical safety, PII protection) are applied automatically from study configuration. Provider-configurable layers (monitoring disclosure, persona, interaction context) can be customized, left at platform defaults, or explicitly disabled. Per-conversation layers (domain guidance, custom instructions, safety-plan context) are added when set. The assembled layers are concatenated in order into a single system prompt and sent to the LLM.}
  \label{fig:prompt}
\end{figure}

\begin{itemize}
  \item \textbf{A layered, composable system prompt.} A central tension in flexible human-AI interaction research in mental health is that different studies need fundamentally different AI behavior, while all must share similar safety commitments. A monolithic prompt that suits a clinical study will break a pedagogical exercise, and vice versa. Cloze resolves this by assembling the system prompt at conversation time from ordered layers with differing editability (Figure~\ref{fig:prompt}). Universal layers (forbidden content, crisis protocols, honesty requirements, persona guardrails) cannot be removed by providers. Conditional study-context layers (clinical restrictions, personal information protection) are applied automatically from study configuration. Provider-configurable layers cover persona, interaction context, and monitoring disclosure, and per-conversation layers carry domain guidance, the study's own protocol, and personalized safety-plan context. The same deployment can therefore run a depression study and a cognitive-skills exercise with identical safety floors and entirely different interaction patterns, and the configuration is explicit and auditable rather than hidden in bespoke prompts.
\end{itemize}

\begin{itemize}
  \item \textbf{Provider abstraction for portability and comparison.} A unified model interface spans OpenAI, Anthropic, Google, and local models served through Ollama, with per-model generation settings such as temperature and token limits. This enables within-cohort cross-model comparison, per-phase model selection, and provider fallback, and allows the same study design to run against a cloud-hosted frontier model or a local open-weight model without modification.

  \item \textbf{Safety as defense-in-depth.} Safety is enforced at the prompt layer through versioned, on-disk constitutional prompts, at the LLM layer through provider safety settings, retry and backoff handling, and human-readable error states, and at the application layer through consent modals, turn limits, and Stanley-Brown structured safety plans with per-participant anti-pattern constraints. Prompt-based safety nonetheless has a limit: provider-configurable layers sit beneath the universal layers, so enforcement ultimately depends on the model respecting instruction ordering. We mitigate this through a provider-trust model in which configuration is restricted to authenticated researchers, together with audit logging, and we do not claim guarantees.
\end{itemize}

Two further design choices matter for research use: \textbf{study flows} (always-available, phased, and recurring) express the longitudinal structure directly, and \textbf{local/on-premises deployment} lets the entire stack, including the model, run within an institution's own infrastructure for data-residency-constrained protocols.

\section*{Research impact}

Cloze has been developed following discussions with lived experience experts and clinical professionals, incorporating the different points of view to create a platform that aligns with both patients' and healthcare professionals' needs. Cloze has been demonstrated at institutions like Oxford University and is operational and in use today with active IRBs, allowing its use in care settings. The platform currently supports a cognitive-skills-training study at Beth Israel Deaconess Medical Center that guides participants through cognitive-restructuring practice via a dual-persona, protocol-defined exercise. Alpha evaluation is being conducted deliberately on a pedagogical application that makes no therapy claim. The study exercises the platform's core flexibility. It runs with the clinical safety layer and individual safety plans disabled, and supports the research protocols around therapy homework engagement, while the universal crisis, forbidden-content, and persona-guardrail floor stays active beneath it. The layered prompt composer, multi-provider model interface, study flows, safety planning, keyword-based safety alerting (Cloze-Guard v0), de-identified participant management, and report generation are all implemented and in use.

Cloze is released under AGPL-3.0; upon completion of alpha testing, self-deployment documentation, prompt configuration references, and IRB template language will follow. These materials will lower the cost for other groups to run their own controlled human-AI studies. Participant accounts are created with randomized identifiers by default, and the platform collects no direct identifiers as part of participant management. Additionally, one-provider-per-participant isolation is enforced at the database level. Cloze is also designed to integrate well with other open-source research software platforms, such as mindLAMP \cite{vaidyam2022} and REDCap \cite{harris2009}. The ability to locally host the platform alongside other research software enables shared data pipelines and analytics tooling, opening additional research avenues that pair human-AI interaction with real-world contextual data.

Alpha study teams have requested extensions that go beyond this core platform, and we are building them as enhancements. A trained LLM-based safety classifier for Cloze-Guard, evaluated against established crisis-detection benchmarks, is already built and will complement the shipped keyword-based alerting. Our team's work on AI safety through the MindBench.ai project informs these efforts \cite{dwyer2025} and ensures we have ongoing efficacy evaluation and direct feedback from all stakeholders, including people with mental illness. An expanded in-platform NLP reporting pipeline will operate at the conversation, phase, flow, and cohort levels, alongside structured data export, an analysis API, and real-time crisis alerting. Deeper integration with other open source research platforms like mindLAMP is also underway, with the goal of allowing Cloze chats to benefit from the deeper contextual awareness of participants driven by anonymized data from external streams like smartphones and wearables \cite{flathers2025}.

It is worth reiterating that the goal of our platform is not to offer a product but rather a tool for more rigorous and reproducible science. The inherent capabilities of AI models are now expanding exponentially, and this precludes the ability, let alone the value, of single-use-case customized AI research platforms. By enabling research teams to safely and reproducibly harness the newest and safest AI models, we hope that AI clinical research can also progress as rapidly and safely as we see the broader AI space today.

\section*{Acknowledgements}

Initial development of Cloze was supported by a grant from the Cosmos Institute to F.C. Development of Cloze's mindLAMP integration tooling was supported by the MEXA Accelerator program, funded by Wellcome Trust and administered by Neuromatch.

\end{document}